\documentclass[12pt]{iopart}

\usepackage{epsf}
\usepackage{amssymb}
\usepackage{amsfonts}
\usepackage{graphicx}
\usepackage{graphics}
\usepackage{epstopdf}
\usepackage{cite}

\newcommand{\xt}{x_{_\perp}}
\newcommand{\ptrans}{p_{_\perp}}

\newcommand{\alphas}{\alpha_s}

\newcommand{\sqrtsnn}{\sqrt{s_{_{\mathrm{NN}}}}}
\newcommand{\dd}{{\rm d}}
\newcommand{\X}{{\rm X}\,}
\newcommand{\A}{{\rm A}}

\newcommand{\rag}{R^{^\A}_{_G}}
\newcommand{\rafd}{R^{^\A}_{_{F_2}}}

\newcommand{\ga}{G^{^\A}}
\newcommand{\gp}{G^{^p}}

\newcommand{\rpa}{R_{_{p\A}}}
\newcommand{\rda}{R_{_{d\A}}}

\newcommand{\rdau}{R_{_{d {\rm Au}}}}
\newcommand{\gtarg}{G^{^{\A}}}
\newcommand{\gproj}{G^{^p}}
\newcommand{\ftarg}{F^{^{\A}}}
\newcommand{\fproj}{F^{^p}}
\def\cO#1{{{\cal{O}}}\left(#1\right)}

\begin{document}

\title[Probing gluon shadowing with forward photons at RHIC]{Probing gluon shadowing with forward photons at RHIC}

\author{Fran\c{c}ois Arleo \footnote{Talk given at Quark Matter 2008, Jaipur (India), February 4-10, 2008.}}
\address{LAPTH, UMR5108, Universit\'e de Savoie, CNRS,\\ BP 110, 74941 Annecy-le-Vieux cedex, France}
\ead{arleo@lapp.in2p3.fr}

\author{Thierry Gousset}
\address{Subatech, UMR 6457, Universit\'e de Nantes,\\ Ecole des Mines de Nantes, IN2P3/CNRS.\\
 4 rue Alfred Kastler, 44307 Nantes cedex 3, France}
 \ead{gousset@subatech.in2p3.fr}

\begin{abstract}
There is a major need to better constrain nuclear parton densities in order to provide reliable perturbative QCD predictions at the LHC as well as to probe possible non-linear evolution at small values of $x$. In these proceedings, we discuss how the production of prompt photons at large rapidity in $p$--$p$ and $d$--Au collisions at RHIC ($\sqrtsnn=200$~GeV) is sensitive to the nuclear modifications of gluon distributions at $x=\cO{10^{-3}}$ and at rather low scales, $Q^2\sim$~10~GeV$^2$. The nuclear production ratio, $\rda=\sigma(d+\A\to\gamma+\X)/(2A\times\sigma(p+p\to\gamma+\X))$, is computed for isolated prompt photons at NLO using the nDSg nuclear parton densities, in order to assess the visibility of the signal. We also emphasise that the expected counting rates in a year of running at RHIC are large, indicating that $\rda$ could be measured with a high statistical accuracy.
\end{abstract}

\section{Gluons from photons}

An important experimental and theoretical effort has been pursued over the last fifteen years aiming at constraining the gluon density in a proton, $\gp$. The current precision on $\gp(x,Q^2)$ has reached a few percent level at not too large nor not too small values of Bjorken-$x$ , $10^{-4}\lesssim x \lesssim 10^{-1}$, and at large virtualities, $Q^2\gtrsim 10$~GeV$^2$. Lacking a similar facility such as HERA using heavy ion beams, the gluon density in a nucleus, $\ga(x,Q^2)$~(see~e.g.~\cite{Armesto:2006ph}) --~or equivalently the normalised nuclear gluon distribution ratio $\rag(x,Q^2)=\ga(x,Q^2)\big/A\ \gp(x,Q^2)$~--
is determined with a poor accuracy, say $\sim 20\%$, and only for rather large values of $x$, $x\gtrsim 10^{-2}$. As a consequence, the precision level for perturbative QCD cross sections in nuclear collisions ($p$--A or A--A) presently taken at RHIC or soon to be taken at the LHC is far from that achieved in $p$--$\bar{p}$ and $p$--$p$ collisions at Tevatron and LHC. Let us also mention that probing accurately $\ga(x, Q^2)$ is also interesting in itself in order to shed light on non-linear evolution expected in QCD at small $x$; a field which has recently received a lot of attention (see e.g.~\cite{Gelis:2007kn}). For these reasons, any way leading to a better control of the gluon nuclear density needs to be considered and properly quantified. Based on a recent study~\cite{Arleo:2007js}, we explored in this talk how the production of prompt photons in hadron--nucleus reactions at RHIC and LHC may serve as a sensitive probe of $\rag(x,Q^2)$.

At leading-order (LO) in $\alphas$, large-$\ptrans$ prompt photon production proceeds either via the annihilation process, $q\bar{q}\to{g}\gamma$, or by Compton scattering, $q(\bar{q})g\to{q}(\bar{q})\gamma$. Note however that, below $\xt\equiv2\ptrans/\sqrt{s}<0.1$, the former channel is less than $10\%$ of the latter. For clarity purposes, let us consider it to be negligible in the following. In the collinear factorisation framework, the inclusive production cross section in $p$--A collisions can simply be expressed as a product of parton distribution functions\footnote{Parton distributions need to be evaluated at a factorisation scale $\cO{\ptrans}$. This scale dependence does not play an important role in the present discussion and is therefore not made explicit in the following.} (PDF), in the proton and in the nucleus, times the partonic cross section, $\hat{\sigma}$:~\cite{Arleo:2007js}
\begin{eqnarray}\label{eq:qcdloxs}
\frac{\dd^3\sigma(p\ \A\to\gamma\ \X)}{\dd{y}\ \dd^2\ptrans}
= \int_{\xt e^y/2}^{1-\xt e^{-y}/2}
 \dd{v} &&\Big[\fproj\left(\frac{\xt e^y}{2v}\right)
\gtarg\left(\frac{\xt e^{-y}}{2(1-v)}\right)\,\hat{\sigma}(v)\\\nonumber
&&+\gproj\left(\frac{\xt e^y}{2v}\right)
\ftarg\left(\frac{\xt e^{-y}}{2(1-v)}\right)\,\hat{\sigma}(1-v)\Big]
\end{eqnarray}
The sum over the quark flavours is implicit in the definition of $F(x)\equiv F_2(x)/x$. The integral over $v$, stemming from the rapidity of the unobserved gluon recoiling jet, $y_{_{\rm jet}}=$asinh$[v^{-1}-(1-v)^{-1}]$, therefore spoils the simple relationship between the observable on the one hand, and the nuclear parton densities --~which we would like to constrain~-- on the other hand.
However, as noticed in~\cite{Arleo:2007js}, the fairly slow variation of $\rag(x)$ and $\rafd(x)$ with $x$ as compared to that of the PDFs themselves allows Eq.~(\ref{eq:qcdloxs}) to be approximated by:
\begin{eqnarray*}\label{eq:qcdloxsapprox}
\frac{\dd^3\sigma(p\ \A\to\gamma\ \X)}{\dd{y}\ \dd^2\ptrans}
&\simeq& \rag(\xt e^{-y})\ \int
 \dd{v}\  \fproj\left(\frac{\xt e^y}{2v}\right)
\gproj\left(\frac{\xt e^{-y}}{2(1-v)}\right)\,\hat{\sigma}(v)\\\nonumber
&+&\rafd(\xt e^{-y}) \int
 \dd{v}\  \gproj\left(\frac{\xt e^y}{2v}\right)
\fproj\left(\frac{\xt e^{-y}}{2(1-v)}\right)\,\hat{\sigma}(1-v)\\
&\simeq& \rag(\xt e^{-y})\ I_{qg}(y) + \rafd(\xt e^{-y})\ I_{gq}(y),
\end{eqnarray*}
where we used the fact that the integral is dominated by $v\simeq 1/2$. Within this approximation, the inclusive prompt photon cross section is simply a linear combination of $\rag$ and $\rafd$, both being evaluated at $\xt e^{-y}$. The above relationship simplifies even further in three kinematic limits. At mid-rapidity, $I_{qg}=I_{gq}$, while we have $I_{qg}(y)\gg I_{gq}(y)$ at forward rapidity, and $I_{qg}(y)\ll I_{gq}(y)$ at backward rapidity. The last two inequalities arise from the dynamics of partons inside the proton: large-$x$ is dominated by valence while gluons are dominant at small $x$. Therefore, photons produced at large positive (respectively, negative) rapidity necessarily originate from the scattering of a valence quark inside the proton (nucleus) and from a gluon inside the nucleus (proton). Defining the nuclear production ratio as
\begin{equation*}
\label{eq:ratio}
  \rpa(\xt) = \frac{1}{A} \ \
  \frac{\dd^3\sigma}{\dd{y}\ \dd^2\ptrans}(p+\A\to\gamma+\X)
  \Big/ \frac{\dd^3\sigma}{\dd{y}\ \dd^2\ptrans}(p+p\to\gamma+\X),
\end{equation*} 
it immediately follows that:\\[0.3cm]
  \begin{tabular}{p{5.cm} p{10.cm}}
    $\bullet\,$ at mid-rapidity:  & $\rpa(\xt,y=0)\ \simeq\ 0.5\ \rafd(\xt)+ 0.5\ \rag(\xt)$; \\[0.3cm]
    $\bullet\,$ at forward rapidity:    & $\rpa(\xt e^{-y})\ \simeq\ \rag(\xt)$;\\[0.3cm]
    $\bullet\,$ at backward rapidity: & $\rpa(\xt e^{-y})\ \simeq\ \rafd(\xt)$.
  \end{tabular}~\\

We identify at least two reasons which could somehow complicate the above relationships. First of all, photons may also come from the collinear fragmentation of hard quarks and gluons produced in the partonic process. In that case, the photon transverse momentum can no longer be related to the momentum fractions of quarks and gluons inside the incoming hadrons, and hence to the proton/nucleus parton densities. However, a way to get rid of most of these fragmentation photons is to apply isolation criteria. In that sense, the production of {\it isolated} photons is a cleaner probe of $\rag$ and $\rafd$ rather than the inclusive yield. Furthermore, the above approximations have been derived at leading order, which may well be spoiled when higher order corrections are taken into account.

\section{Phenomenology at RHIC}

\begin{figure}[h]
~\hspace{-2.cm}
  \begin{minipage}[t]{9.8cm}
    \begin{center}
      \includegraphics[height=8.cm]{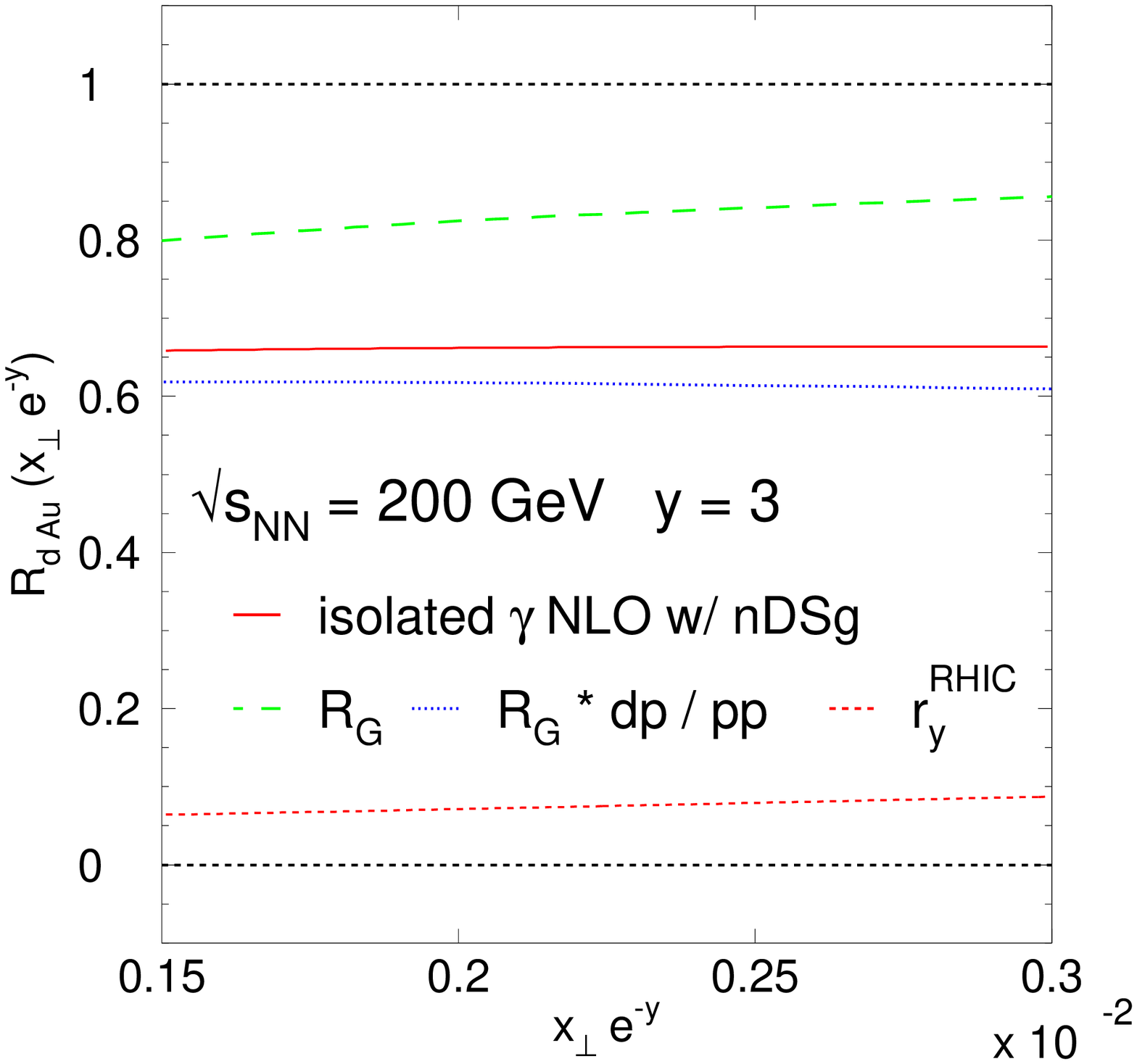}
    \end{center}
    \caption{$\rdau$ of isolated photon production (y=3) at $\sqrtsnn=200$~GeV (see text for details).}
    \label{fig:isoy3rhic}
\end{minipage}
~\hspace{-2.cm}
  \begin{minipage}[t]{9.8cm}
    \begin{center}
      \includegraphics[height=8.cm]{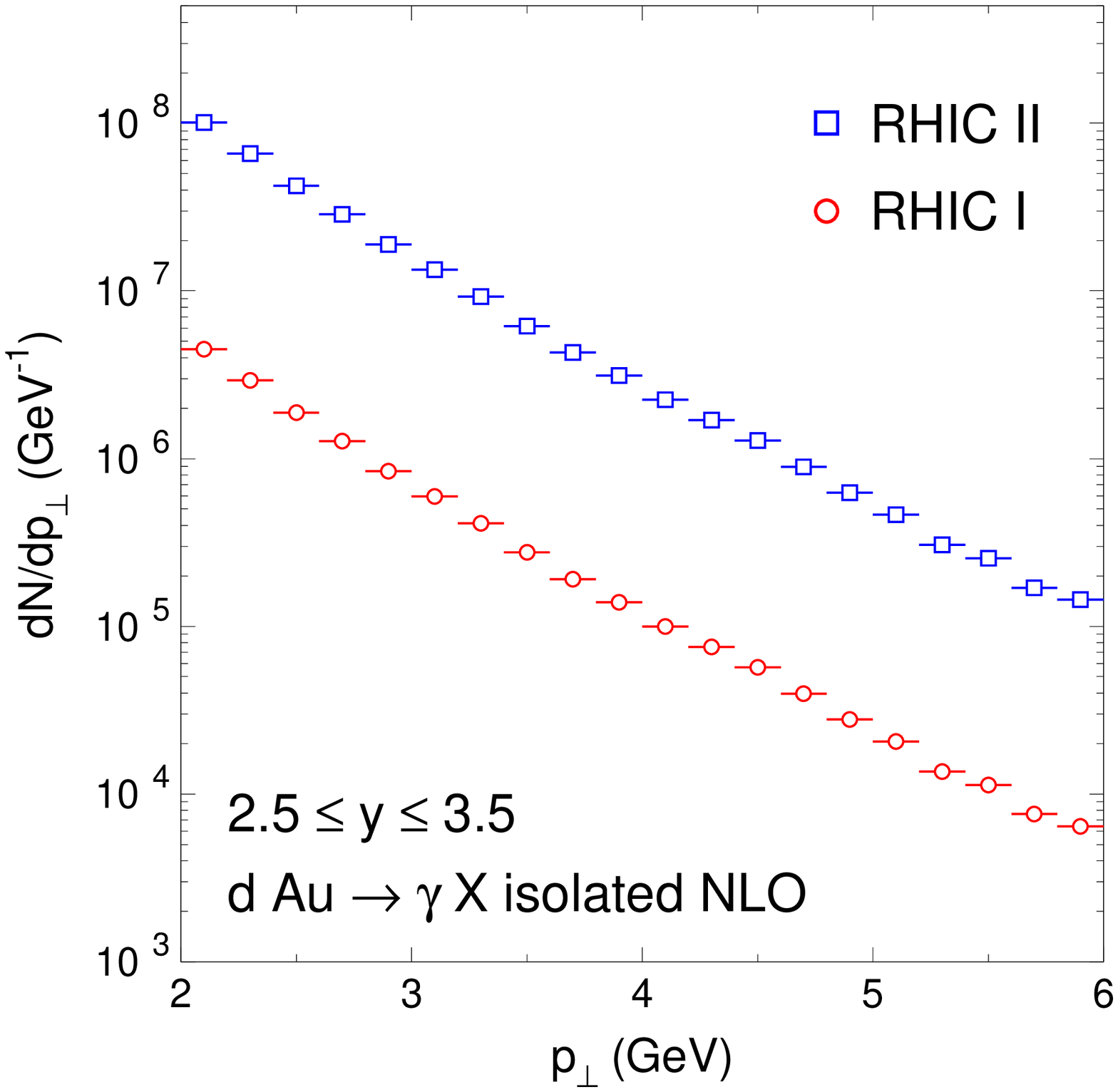}
    \end{center}
    \caption{Counting rates for isolated photons in $d$--Au collisions at $\sqrtsnn=200$~GeV at RHIC~I and RHIC~II.}
    \label{fig:rates}
\end{minipage}
\end{figure}

In order to quantify the accuracy of the above approximations, the nuclear production ratio of isolated photons at forward rapidity (taking $y=3$, a region which should be accessible both by the PHENIX~\cite{OBrien:2006cv} and the STAR~\cite{Bland:2005uu} experiments at RHIC) is computed at NLO (using the calculations of Ref.~\cite{Aurenche:2006vj}) in $d$--Au over $p$--$p$ collisions at $\sqrtsnn=200$~GeV (Fig.~\ref{fig:isoy3rhic}). 

The expected suppression using the nDSg NLO~\cite{deFlorian:2003qf} nuclear parton densities (solid line) is $\rpa\simeq 0.65$, that is much below $\rag\simeq 0.8$ (dashed line). The main reason for such a mismatch is, however, coming from an isospin effect, as we compare systems which projectiles differ: a proton and a deuteron. Recall indeed that, being electromagnetic probes, photons couple more to $u$ quarks than to $d$ quarks ($e_u^2=4e_d^2$). Therefore, since the deuteron is poorer in $u$ than the proton is, prompt photon production is suppressed in $d$--$p$ with respect to $p$-$p$ scattering. The proton and neutron seas being roughly symmetric, the suppression should be visible as long as valence quarks are involved, that is at large $x_1\gtrsim 0.1$ which is the case at RHIC at forward rapidity. For the sake of a rough estimate, let us assume that only scattering of valence quarks with gluons is involved at large $x$ and that $u^p\simeq 2d^p$, then the isospin-corrected relationship we could expect between $\rda$ and $\rag$ is:
$\rda(x)\ \simeq\ (5u+5d)/(8u+2d)\times\rag(x) \simeq 5/6\times\rag(x)$ (in practice the isospin correction factor is given by $R_{_{dp}}=\sigma(dp)/\sigma(pp)$, which is under better theoretical control). As can be seen in Fig.~\ref{fig:isoy3rhic} (dotted line) the isospin-corrected $\rag$ is in pretty good agreement, say within $10\%$, with the predicted $\rda$ (see also the ratio of these two curves, minus one, $r_y^{\rm RHIC}$, bottom line).

From this analysis~\cite{Arleo:2007js}, it appears that photon production is a sensitive probe of the nuclear modifications of parton densities. This observable is, nevertheless, only interesting if the nuclear production ratio can be measured with a high statistical accuracy, that is if reasonable rates can be achieved. Using an integrated luminosity of ${\cal L}=0.02$~pb$^{-1}$ and ${\cal L}=0.45$~pb$^{-1}$ at RHIC~I and RHIC~II~\cite{ncctdr}, respectively, the expected NLO rates are plotted in Fig.~\ref{fig:rates} as a function of $\ptrans$ in the $y=2.5$--$3.5$ rapidity range. Fortunately, the photon yield is pretty large up to rather high $\ptrans$ and already at RHIC~I, which is very encouraging. Provided the experimental systematic uncertainties remain low, there is a real hope that the suppression of isolated photon production soon to be measured at RHIC will allow for tight constraints to be set on $G^{\rm A}/G^p$ at rather small values of $x$.

~\\[-0.8cm]
\providecommand{\newblock}{}

\end{document}